# Change of carrier density at the pseudogap critical point of a cuprate superconductor

S. Badoux[1], W. Tabis[2,3], F. Laliberté[2], G. Grissonnanche[1], B. Vignolle[2], D. Vignolles[2], J. Béard[2], D. A. Bonn[4,5], W. N. Hardy[4,5], R. Liang[4,5], N. Doiron-Leyraud[1], Louis Taillefer[1,5] and Cyril Proust[2,5]

*1 Département de physique & RQMP, Université de Sherbrooke, Sherbrooke, Québec J1K 2R1, Canada*

*2 Laboratoire National des Champs Magnétiques Intenses (CNRS, INSA, UJF, UPS), Toulouse 31400, France*

*3 AGH University of Science and Technology, Faculty of Physics and Applied Computer Science, 30-059 Krakow, Poland*

*4 Department of Physics & Astronomy, University of British Columbia, Vancouver, British Columbia V6T 1Z1, Canada*

*5 Canadian Institute for Advanced Research, Toronto, Ontario M5G 1Z8, Canada*

**The pseudogap is a central puzzle of cuprate superconductors. Its connection to the Mott insulator at low doping $p$ remains ambiguous[1] and its relation to the charge order[2,3,4] that reconstructs the Fermi surface[5,6] at intermediate $p$ is still unclear[7,8,9,10]. Here we use measurements of the Hall coefficient in magnetic fields up to 88 T to show that Fermi-surface reconstruction by charge order in YBa$_2$Cu$_3$O$_y$ ends sharply at a critical doping $p = 0.16$, distinctly lower than the pseudogap critical point at $p^* = 0.19$ (ref. 11). This shows that pseudogap and charge order are separate phenomena. We then find that the change of carrier density from $n = 1 + p$ in the conventional metal at high $p$ (ref. 12) to $n = p$ at low $p$ (ref. 13) – a signature of the lightly doped cuprates – starts at $p^*$. This shows that**

**pseudogap and antiferromagnetic Mott insulator are linked.**

Electrons in cuprate materials go from a fairly conventional metallic state at high hole concentration (doping) $p$ to a Mott insulator at $p = 0$. How the system evolves from one state to the other remains a fundamental question. At high doping, the Fermi surface of cuprates is well established. It is a large hole-like cylinder whose volume yields a carrier density $n = 1 + p$, as measured, for example, by angle-resolved photoemission spectroscopy (ARPES)[14], in agreement with band structure calculations. The carrier density can also be measured using the Hall coefficient $R_H$ since in the limit of $T = 0$ the Hall number $n_H$ of a single-band metal is such that $n_H = n$. Indeed, in the cuprate $Tl_2Ba_2CuO_{6+\delta}$ (Tl-2201), the normal-state Hall coefficient $R_H$ at $p \sim 0.3$, measured at $T \rightarrow 0$ in magnetic fields large enough to suppress superconductivity, is such that $n_H = V / e\, R_H = 1 + p$, where $e$ is the electron charge and $V$ the volume per Cu atom in the $CuO_2$ planes[12,15].

By contrast, at low doping, measurements of $R_H$ in $La_{2-x}Sr_xCuO_{4+\delta}$ (LSCO) (ref. 13) and $YBa_2Cu_3O_y$ (YBCO) (ref. 16) yield $n_H \sim p$, below $p \sim 0.08$. Having a carrier density equal to the hole concentration, $n = p$, is known to be an experimental signature of the lightly doped cuprates. The question is: At what doping does the transition between those two limiting regimes take place? Specifically, does the transition from $n = 1 + p$ to $n = p$ occur at $p^*$, the critical doping for the onset of the pseudogap phase? The pseudogap is a depletion of the normal-state density of states that appears below $p^* \sim 0.19$ (ref. 11), whose origin is a central puzzle in the physics of correlated electrons and the subject of much debate.

To answer this question using Hall measurements, one needs to reach low temperatures, which requires the use of large magnetic fields to suppress superconductivity. The only prior high-field study of cuprates that goes across $p^*$ was performed on LSCO (ref. 17), a cuprate superconductor with a relatively low critical temperature ($T_c < 40$ K) and critical field ($H_{c2} < 60$ T). For mainly two reasons, studies on LSCO were inconclusive on the transition from $n = 1 + p$ to $n = p$. First, the Fermi surface of overdoped LSCO undergoes a Lifshitz transition from a hole-like to an electron-like surface as its band structure crosses a saddle-point van Hove singularity at $p \sim 0.2$ (ref. 18). This transition causes large changes in $R_H(T)$ (ref. 15) that can mask the effect of the pseudogap onset at $p^* \sim 0.19$. The second problem is the ill-defined impact of the charge-density-wave (CDW) modulations that develop at low temperature in a doping range near $p \sim 0.12$, not only in LSCO (ref. 19), but also in $Bi_2La_{2-x}Sr_xCuO_{6+\delta}$ (Bi-2201) (ref. 20), the other cuprate whose Hall coefficient was measured in high fields[21]. Such CDW modulations should cause a reconstruction of the Fermi surface, and hence change $R_H$ at low temperature[6]. Therefore, the anomalies in $n_H$ vs $p$ observed below 60 K in LSCO and Bi-2201 between $p \sim 0.1$ and $p \sim 0.2$ (refs. 17, 21) are most likely the combined result of three effects that have yet to be disentangled: Lifshitz transition, Fermi-surface reconstruction (FSR) and pseudogap.

Here we turn to YBCO, a cuprate material with several advantages. First, it is one of the cleanest and best ordered of all cuprates, thereby ensuring a homogeneous doping ideal for distinguishing nearby critical points. Second, the location of the pseudogap critical point is well established in YBCO, at $p^* = 0.19 \pm 0.01$ (ref. 11). Third, the Lifshitz transition in YBCO occurs at $p > 0.29$ (ref. 22), well above $p^*$. Fourth, the CDW modulations in YBCO have been thoroughly characterized. They are detected by X-ray diffraction (XRD) between $p \sim 0.08$ and $p \sim 0.16$ (refs. 23, 24), below a temperature $T_{XRD}$ (Fig. 1a). Above a threshold magnetic field, CDW order is detected by NMR (refs. 2, 25), below a temperature $T_{NMR}$ (Fig. 1b). Fifth, the FSR caused by the CDW modulations has a well-defined signature in the Hall effect of YBCO: $R_H(T)$ decreases smoothly to become negative at low temperature[6] – the signature of an electron pocket in the reconstructed Fermi surface. Prior Hall measurements in magnetic fields up to 60 T show that the CDW-induced FSR begins sharply at $p = 0.08$ and persists up to $p = 0.15$, the highest doping reached so far[6].

We have performed Hall measurements in YBCO up to 88 T, allowing us to extend the doping range upwards, and hence track the normal-state properties across $p^*$, down to at least $T = 40$ K. Our complete data on four YBCO samples with dopings $p = 0.16, 0.177, 0.19$ and $0.205$ are displayed in Figs. S1, S2, S3 and S4, respectively. In Fig. 2, we compare field sweeps of $R_H$ vs $H$ at $p = 0.15$ (from ref. 6) and $p = 0.16$, at various temperatures down to 25 K. The difference is striking. At $p = 0.15$, the high-field isotherms $R_H(H)$ drop monotonically with decreasing $T$ until they become negative at low $T$. At $p = 0.16$, $R_H(H)$ never drops. Fig. 3 compares the temperature evolution of the normal-state $R_H$ at different dopings. In Fig. 3a, we see that $R_H(T)$ at $p = 0.16$ shows no sign of the drop to negative values displayed at $p = 0.12, 0.135$ and $0.15$, at least down to $T = 40$ K. Having said this, and although the isotherms at $T = 25$ K and 30 K are consistent with a constant $R_H$ below $T = 50$ K (Fig. 2), we cannot exclude that $R_H(T)$ starts to drop below 40 K. However, even if it did, the onset temperature for FSR would have to be much lower than it is at $p = 0.15$, and it would extrapolate to zero at $p < 0.165$ (Fig. S5). We find that the critical doping above which there is no FSR in the normal state of YBCO at $T = 0$ is $p_{FSR} = 0.16 \pm 0.005$. Because this is in excellent agreement with the maximal doping at which short-range CDW modulations have been detected by XRD, namely $p_{XRD} = 0.16 \pm 0.005$ (ref. 24), and it is consistent with the region of CDW order seen by NMR (ref. 25) (Fig. 1b), we conclude that the critical doping where CDW order ends in YBCO is $p_{CDW} = 0.16 \pm 0.005$.

An onset of CDW order at $p_{CDW} = 0.16$ is distinctly lower than the onset of the pseudogap. Indeed, extensive analysis of the normal-state properties of YBCO above $T_c$ yields $p^* = 0.19 \pm 0.01$ (ref. 11). The critical point $p^*$ can also be located by suppressing superconductivity with 6% Zn impurities[26], which shrinks the $T_c$ dome to a small region centered around $p^* = 0.19$ (Fig. 1a). This robustness of $p^*$ confirms that CDW order and pseudogap are distinct phenomena, since CDW modulations are rapidly weakened by Zn substitution[27]. Applying a field of 50 T produces a small $T_c$ dome peaked at exactly the same doping, showing that in the normal state (whether Zn-induced or field-induced) $p^* = 0.19 \pm 0.01$ (Fig. 1).

We have arrived at our first major finding: pseudogap and CDW order onset at two distinct and well-separated critical dopings. Just as $T_{XRD}$, $T_{NMR} < T^*$ (Fig. 1), we now find that $p_{CDW} < p^*$, in the normal state of YBCO. This contrasts with the simultaneous onset of pseudogap and short-range CDW modulations observed in the zero-field superconducting state of $Bi_2Sr_2CaCuO_{8+x}$ (Bi-2212) by STM (ref. 8).

Having established that the FSR due to CDW order ends at $p_{FSR} = 0.16$, let us see what happens at higher $p$. At $p = 0.205$, the temperature dependence of $R_H$ in YBCO is similar to that of Tl-2201 (refs. 12, 15) at dopings where the Fermi surface is known to be a single large hole-like cylinder with carrier density $n = 1 + p$ (refs. 14, 15) (Fig. S6). In particular, as $T$ increases from zero, $R_H(T)$ rises initially, because of the growth in inelastic scattering, which is anisotropic around the large Fermi surface[15]. This yields a characteristic peak in $R_H(T)$, at $T \sim 100$ K (Fig. S8). Moving to $p = 0.19$, a qualitative change has taken place (Fig. 3c): $R_H(T)$ now shows no sign of a decrease as $T \rightarrow 0$, down to our lowest temperature of 35 K (Fig. S7). The extrapolated $T = 0$ value, $R_H(0)$, doubles upon crossing $p^*$.

Moving to still lower doping, we see that there is also a major *quantitative* change: the magnitude of $R_H$ at low $T$ undergoes a nearly 6-fold increase between $p = 0.205$ and $p = 0.16$ (Fig. 3b), seen directly in the raw data at $T = 50$ K (Fig. 4a). We attribute this huge increase in $R_H$ to a corresponding decrease in carrier density. In other words, states at the Fermi surface are lost and $R_H(T = 0)$ increases. One may argue that for $p < 0.2$ $R_H(T)$ could decrease below 50 K and reach a value at $T = 0$ such that $n_H = 1 + p$ for all dopings down to $p = 0.16$. In this scenario, the peak in $R_H(T)$ at $T$ = 50 K would be due to an anisotropic inelastic scattering that grows rapidly with underdoping[15]. In the Supplementary Material, we show that this mechanism is inconsistent with the measured resistivity of YBCO, which is essentially independent of doping at $T$ = 50 K (Fig. S8).

In Fig. 4b, we plot $n_H$ vs $p$ and discover that in the normal state of YBCO the transition from the conventional metal at high $p$ (where $n_H$ = 1 + $p$) to the lightly doped regime at low $p$ (where $n_H$ = $p$) starts sharply at $p = p^*$, where the pseudogap opens. This is our second major finding. The observed change in $R_H$ by a factor ~ 6 is now understandable, since $(1 + p^*) / p^* = 6.3$. It is important to note that the huge rise in $R_H(0)$ as $p$ is reduced below $p^*$ is the result of a gradual process that begins at high temperature. As seen in Fig. 3d, the order-of-magnitude growth in $R_H$ with decreasing $p$ seen at $T \rightarrow 0$ is also observed at 300 K. Moreover, this growth is monotonic. Those two facts are consistent with the pseudogap phase, whose characteristic temperature $T^*$ rises monotonically with decreasing $p$, to values exceeding 300 K (Fig. 1a). By contrast, CDW modulations cannot be responsible for the enhanced $R_H(T)$, since their onset temperature is non-monotonic and it never exceeds 150 K (Fig. 1a).

In the pseudogap phase, the topology of the $T$ = 0 Fermi surface in the absence of superconductivity and CDW order is unknown. However, because the pseudogap opens at $k = (0, \pm \pi)$ and $(\pm \pi, 0)$, the electronic states at the Fermi level must lie near $k = (\pm \pi/2, \pm \pi/2)$, where the four nodes of the $d$-wave superconducting gap are located. This is indeed what is observed, in the form of nodal Fermi arcs, for example by ARPES in YBCO (ref. 22) and by STM in Bi-2212 (ref. 8), below $p \sim 0.2$. Given that the relation $n_H = p$ extends down to the lowest dopings (Fig. 4b), two scenarios for these nodal states come to mind. One is associated with the antiferromagnetic order, the other is associated with the Mott insulator.

Antiferromagnetic order with a commensurate wavevector $Q = (\pi, \pi)$ – the order that prevails in YBCO below $p = 0.05$ (Fig. 1) – would reconstruct the large Fermi surface into four small hole-like nodal pockets whose total volume would contain $p$ carriers, so that $n_H = p$ (see sketch in Fig. 4b). In electron-doped cuprates, an antiferromagnetic quantum critical point (QCP) is believed to account for the abrupt drop in carrier density detected in the normal-state Hall coefficient[28]. The question is whether in YBCO magnetic order – present at low temperature up to $p \sim 0.08$ in zero field[29] (Fig. 4b) – could extend up to $p^* = 0.19$ when superconductivity is suppressed by a magnetic field of order 100 T. An antiferromagnetic QCP at $p^*$ in YBCO could account for the linear temperature dependence of the resistivity[30] and possibly also the divergent effective mass[9].

In the second scenario, the pseudogap phase is a consequence of strong correlations associated with Mott physics. Numerical solutions of the Hubbard model find nodal Fermi arcs at low doping and intermediate temperatures[31,32]. At $T \to 0$, it has been argued that the Fermi surface could in fact consist of four hole-like nodal pockets[33,34], whose total volume would contain $p$ carriers. These arcs / pockets develop even though translational symmetry is not broken. The question is whether such a Mott-based pseudogap can appear at a doping as high as $p = 0.19$.

The fact that the normal-state carrier density – measured directly in the archetypal cuprate YBCO at low temperature – drops sharply from $n = 1 + p$ to $n = p$ precisely at $p^*$ reveals a robust and fundamental new fact about the pseudogap phase. We expect that a microscopic understanding of this fact will elucidate the enigmatic behavior of electrons in cuprate superconductors.

## Acknowledgements

L.T. thanks the Laboratoire National des Champs Magnétiques Intenses (LNCMI) in Toulouse for their hospitality and LABEX NEXT for their support while this work was being performed. A portion of this work was performed at the LNCMI, which is supported by the French ANR SUPERFIELD, the EMFL, and the LABEX NEXT. R.L., D.A.B. and W.N.H. acknowledge funding from the Natural Sciences and Engineering Research Council of Canada (NSERC). L.T. acknowledges support from the Canadian Institute for Advanced Research (CIFAR) and funding from NSERC, the Fonds de recherche du Québec - Nature et Technologies (FRQNT), the Canada Foundation for Innovation (CFI) and a Canada Research Chair.

## Author contributions

S.B., W.T., F.L., B.V., D.V., J.B. and C.P. performed the transport measurements at the LNCMI. S.B. and N.D.-L. performed the transport measurements at Sherbrooke. G.G. performed the calculations of the transport coefficients. D.A.B., W.N.H. and R.L. prepared the YBCO single crystals at UBC. S.B., L.T. and C.P. wrote the manuscript, in consultation with all authors. L.T. and C.P. co-supervised the project.

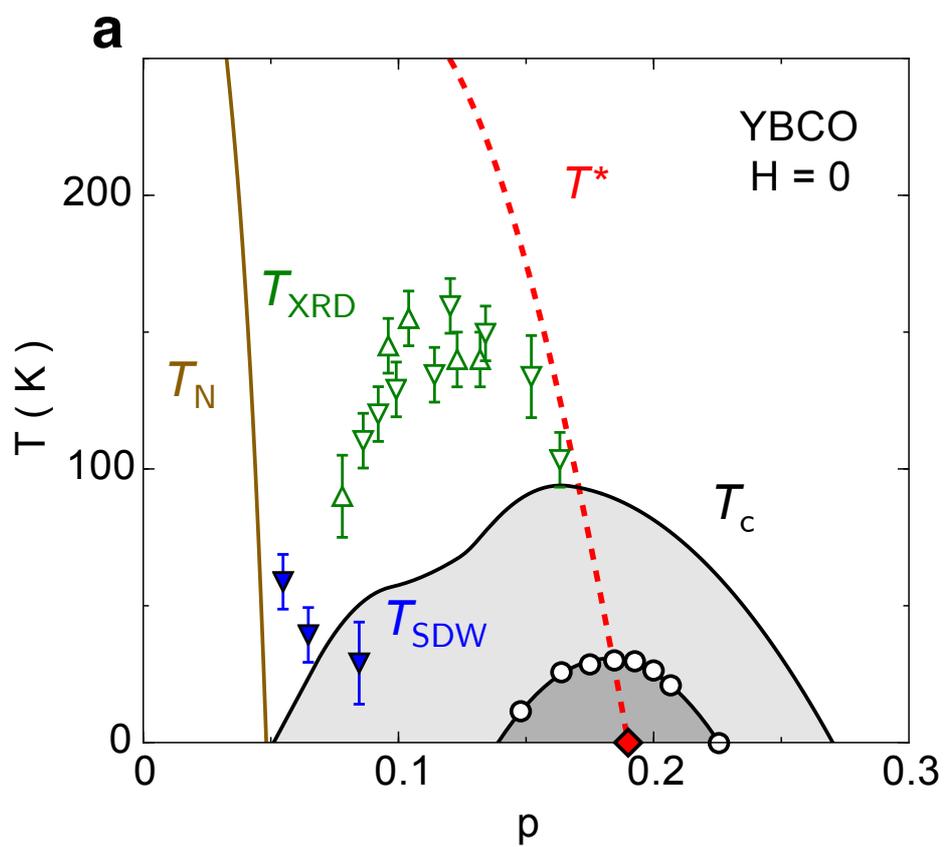

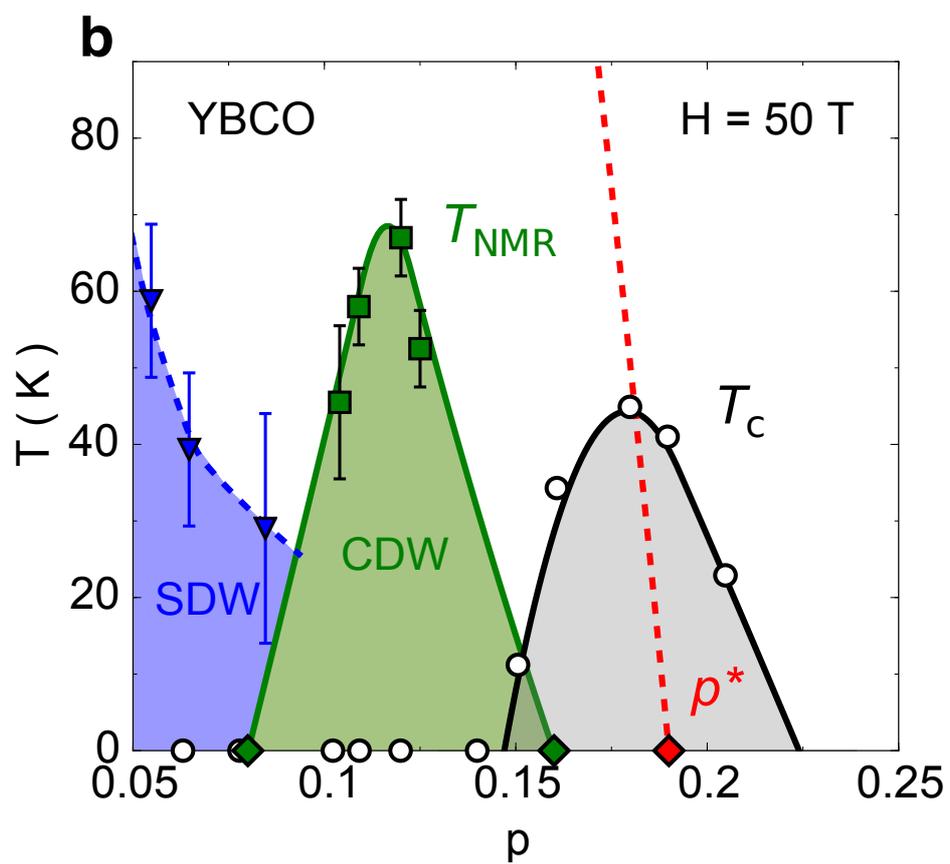

**Fig. 1 | Temperature-doping phase diagram of YBCO.**

**a)** In zero magnetic field ($H = 0$): the superconducting phase (grey dome) lies below $T_c$ (solid black line) and the antiferromagnetic phase lies below $T_N$ (brown line). The small (dark grey) dome shows how $T_c$ is suppressed by substituting 6% of the Cu atoms for Zn (white circles from ref. 26). Short-range charge-density-wave (CDW) modulations are detected by X-ray diffraction below $T_{XRD}$ (up open triangles[23]; down open triangles[24]). Note that the amplitude of the CDW modulations decreases monotonically to zero as doping goes from $p = 0.12$ to $p_{CDW} = 0.16 \pm 0.005$ (ref. 24). Short-range spin-density-wave (SDW) modulations are detected by neutron diffraction below $T_{SDW}$ (blue triangles, from ref. 29). The red dashed line marks the approximate location of the pseudogap temperature $T^*$, while $p^* = 0.19 \pm 0.01$ marks the critical doping below which the pseudogap is known to appear[11] (red diamond). **b)** In a magnetic field $H = 50$ T: above a threshold magnetic field, CDW order is detected by NMR (ref. 2) below a transition temperature $T_{NMR}$ (green squares[25]). The green region is where the Hall coefficient $R_H$ is negative (from ref. 6 and this work). Our Hall data show that Fermi-surface reconstruction, and hence CDW order, ends at $p_{FSR} = 0.16 \pm 0.005$ (green diamond). The fact that the $T_c$ dome at $H = 50$ T (grey) peaks at the same doping as the $T_c$ dome for 6% Zn (panel a) shows that the pseudogap critical point $p^*$ (red diamond) does not move with field. The red dashed line is the same as in panel a. The zero-field SDW phase is reproduced from panel a (blue region).

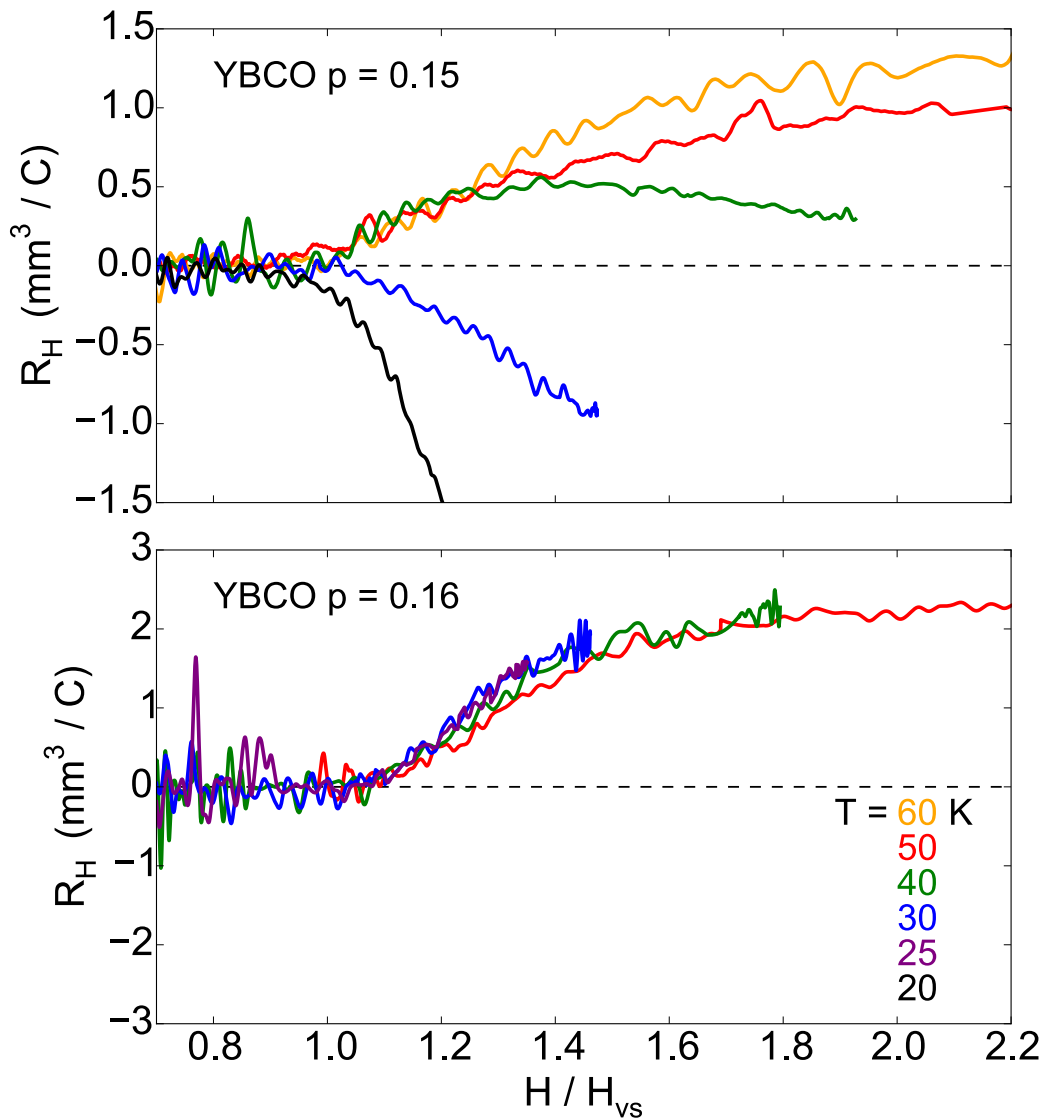

**Fig. 2 | Field dependence of the Hall coefficient in YBCO.**

Hall coefficient of YBCO at various fixed temperatures, as indicated, plotted as $R_H$ vs $H / H_{vs}$, where $H_{vs}(T)$ is the vortex-lattice melting field above which $R_H$ becomes non-zero, for two dopings: $p$ = 0.15 (top panel) and $p$ = 0.16 (bottom panel). Upon cooling, we see that $R_H$ decreases and eventually becomes negative at $p$ = 0.15, while it never drops at $p$ = 0.16.

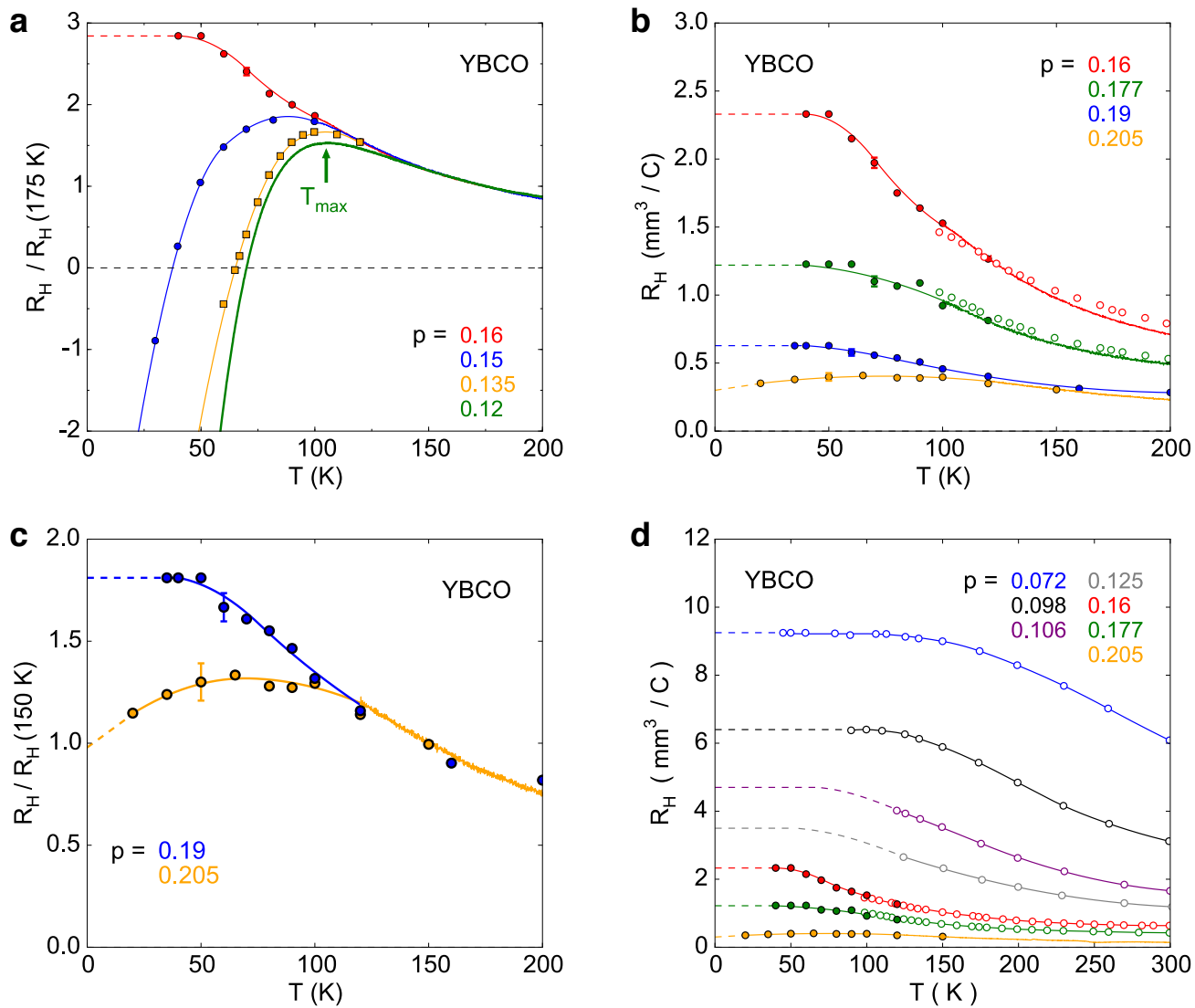

**Fig. 3 | Temperature dependence of the Hall coefficient in YBCO.**

Temperature dependence of the normal-state Hall coefficient of YBCO at various dopings, as indicated. **a)** $R_H$ is normalized by its value at $T$ = 175 K. The solid red, blue, yellow and green curves are temperature sweeps at $H$ = 16 T, above $T$ = 100, 100, 120 and 60 K, respectively. Color-coded lines below those temperatures are a guide through the data points; the red dashed line is a flat extrapolation below 40 K. The data points for $p$ = 0.16 (red) are taken at (or extrapolated to) $H$ = 80 T, from the $R_H$ vs $H$ isotherms in Fig. S1. The data points for $p$ = 0.15 (blue) and $p$ = 0.135 (yellow) are taken from ref. 6 (at $H$ = 55 T). The arrow marks the location of the peak in $R_H$ vs $T$, for $p$ = 0.12 ($T_{max}$). The drop in $R_H(T)$ at low temperature is the signature of Fermi-surface reconstruction (FSR), caused by charge-density-wave (CDW) order. At $p$ = 0.16, no such drop occurs, at least down to 40 K. This reveals that the critical doping for the end of FSR in the doping phase diagram (Fig. 1b) is $p_{FSR}$ = 0.16 ± 0.005 (Fig. S5). **b)** $R_H$ vs $T$ at $p$ = 0.16 and higher, measured at (or extrapolated to) $H$ = 80 T (full circles), from isotherms in Figs. S1, S2, S3 and S4. The solid red, green and yellow curves are temperature sweeps at $H$ = 16 T, above $T$ = 100, 100 and 120 K, respectively. Color-coded lines below those temperatures are a guide to the eye through the data points. The dashed lines are a linear extrapolation below the lowest data point. Open circles are low-field data from ref. 16 for the normal-state $R_H(T)$ of YBCO above $T_c$, for $p$ = 0.16 ($y$ = 6.95, $T_c$ = 93 K) and $p$ = 0.178 ($y$ = 7.00, $T_c$ = 91 K). These data are in excellent quantitative agreement with our own data. **c)** Same as in b), showing the two highest dopings only, with $R_H$ normalized at $T$ = 150 K. The curve at $p$ = 0.19 is qualitatively different from the curve at $p$ = 0.205, showing no sign of a drop at low $T$ (Fig. S7). We attribute the two-fold increase in the magnitude of $R_H$ at $T \to 0$ to a decrease in carrier density as the pseudogap opens at $p^*$, with $p^*$ located between $p$ = 0.205 and $p$ = 0.19. The error bars reflect the relative uncertainty in determining the change in $R_H$ vs $T$ for a given doping. **d)** Same as in b), over a wider range of doping and temperature. For the three curves in the interval 0.09 < $p$ < 0.15, the dashed lines show how the normal-state $R_H(T)$ might extend down to $T$ = 0 in the absence of the FSR caused by CDW order.

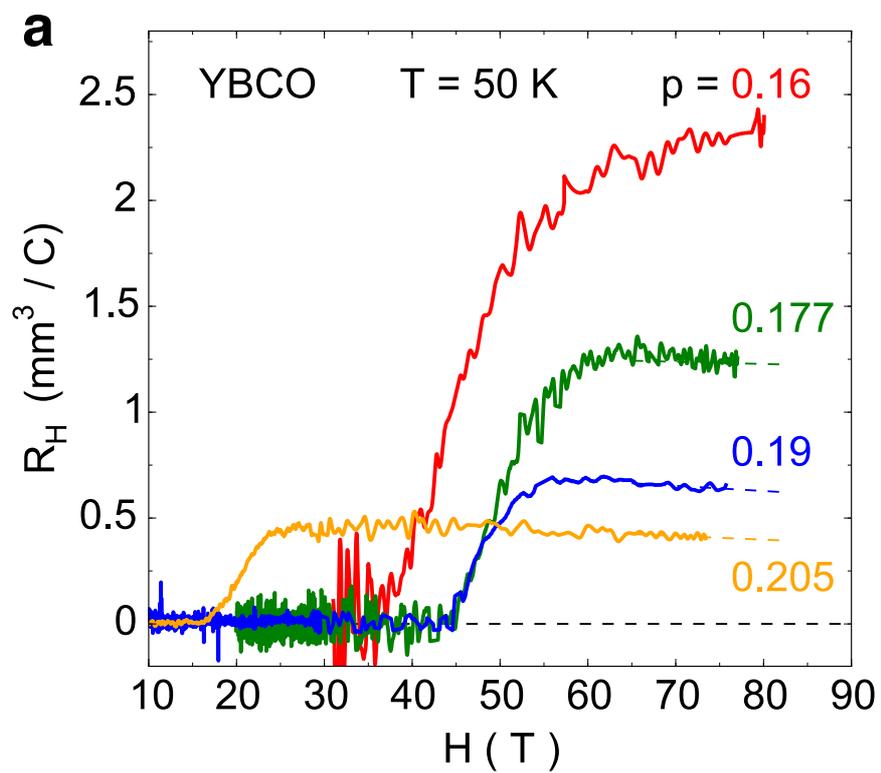
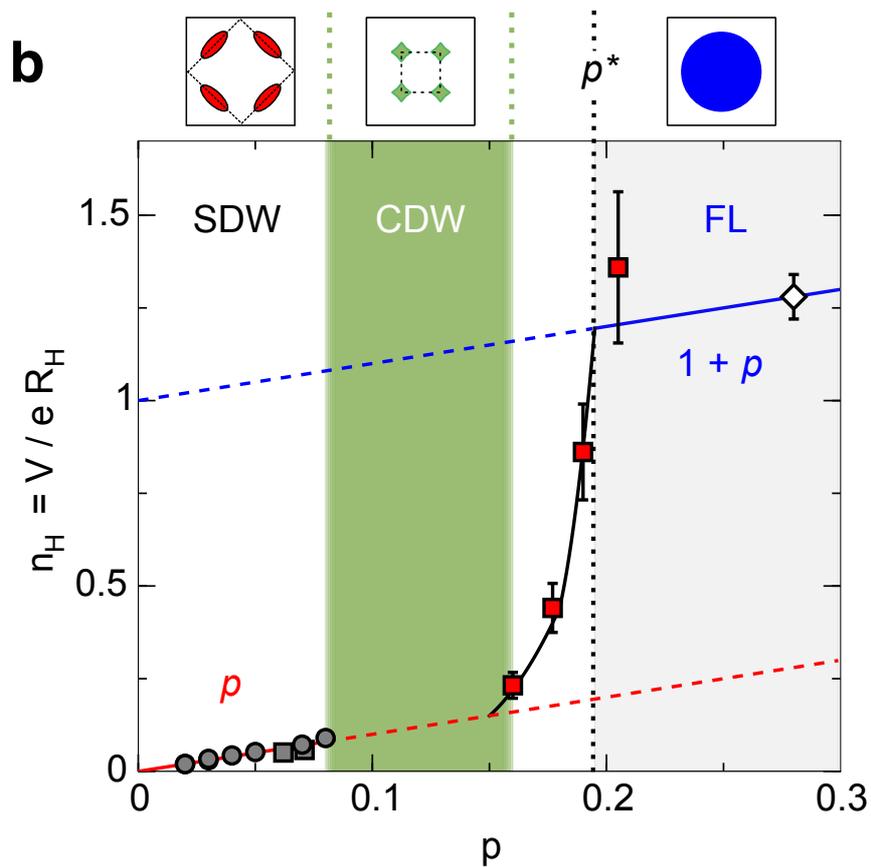

**Fig. 4 | Doping evolution of the normal-state carrier density.**

**a)** Isotherms of $R_H$ vs $H$ in YBCO at $p$ = 0.16, 0.177, 0.19, and 0.205, measured at $T$ = 50 K. Note the huge increase in the value of $R_H$ at $H$ = 80 T (or extrapolated to $H$ = 80 T; dashed lines), by a factor 5.7, when going from $p$ = 0.205 to $p$ = 0.16. **b)** Doping dependence of the Hall number, $n_H = V / e R_H$, a direct measure of the carrier density in hole-doped cuprates, measured in the normal state at $T$ = 50 K for LSCO (circles, ref. 13) and YBCO ($p$ < 0.08, grey squares, ref. 16). For YBCO at $p$ > 0.15 (red squares), we use $R_H$ at $H$ = 80 T from panel a). The white diamond is obtained from the $T$ = 0 limit of $R_H(T)$ in strongly overdoped Tl-2201 (ref. 12). The black line is a guide to the eye. The red curve is $n_H = p$ ; the blue curve is $n_H = 1 + p$. The region where Fermi-surface reconstruction due to CDW order occurs in YBCO is marked as a green band; in that band, $R_H$ < 0. With decreasing $p$, the carrier density is seen to drop rapidly from 1 + $p$ to $p$ at $p^*$ = 0.19 ± 0.01 (black dotted line), the critical doping for the onset of the pseudogap in YBCO (ref. 11; Fig. 1). The error bars (± 15 %) given for our 4 samples (red squares) reflect the uncertainty on the absolute value of $R_H$ (see Supplementary Information). The icons above the figure show a sketch of the normal-state Fermi surface in three of the four doping regions: small nodal hole pockets (red) below $p$ = 0.08, where magnetic order (SDW) prevails at low temperature (Fig. 1b); small electron pockets (green) between $p$ = 0.08 and $p$ = 0.16, where charge order (CDW) prevails at low temperature (Fig. 1b); a single large hole surface (blue) above $p^*$, where the non-superconducting ground state is a Fermi liquid (FL, grey region).

# SUPPLEMENTARY INFORMATION

**Supplementary Figure S1**

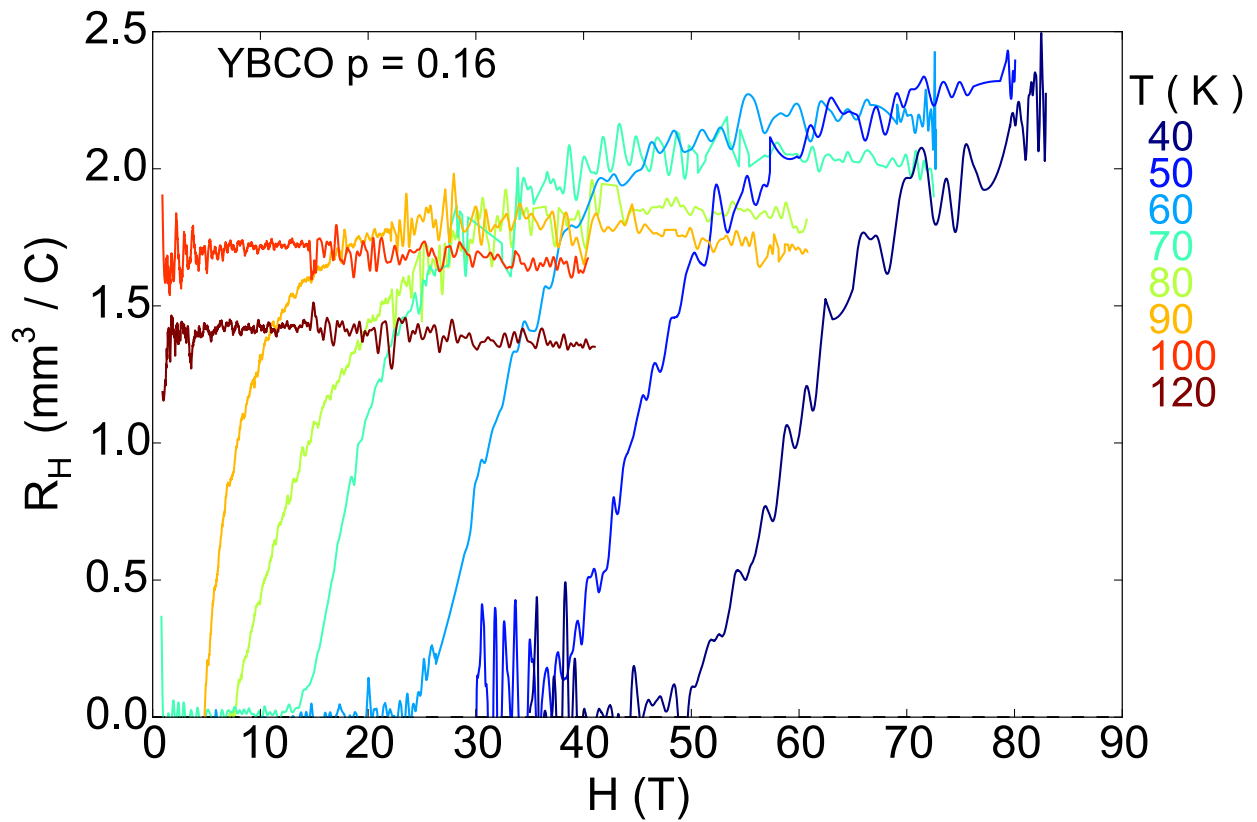

**Figure S1 | Isotherms of $R_H$ vs $H$ in YBCO at $p$ = 0.16.**

Magnetic field dependence of the Hall coefficient $R_H$ in our YBCO sample with $y$ = 6.92 ($T_c$ = 93.5 K; $p$ = 0.161), at various temperatures as indicated.

**Supplementary Figure S2**

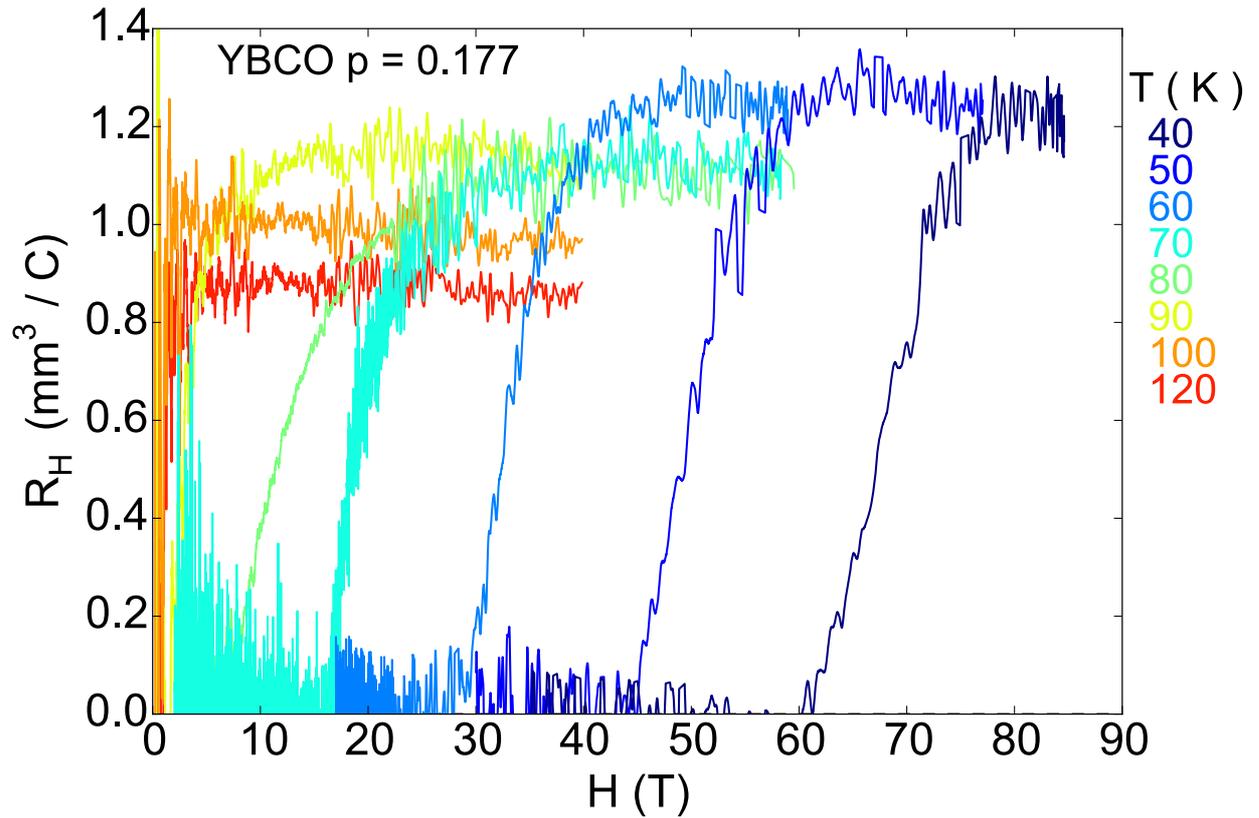

**Figure S2 | Isotherms of $R_H$ vs $H$ in YBCO at $p$ = 0.177.**

Magnetic field dependence of the Hall coefficient $R_H$ in our YBCO sample with $y$ = 6.97 ($T_c$ = 91 K; $p$ = 0.177), at various temperatures as indicated.

**Supplementary Figure S3**

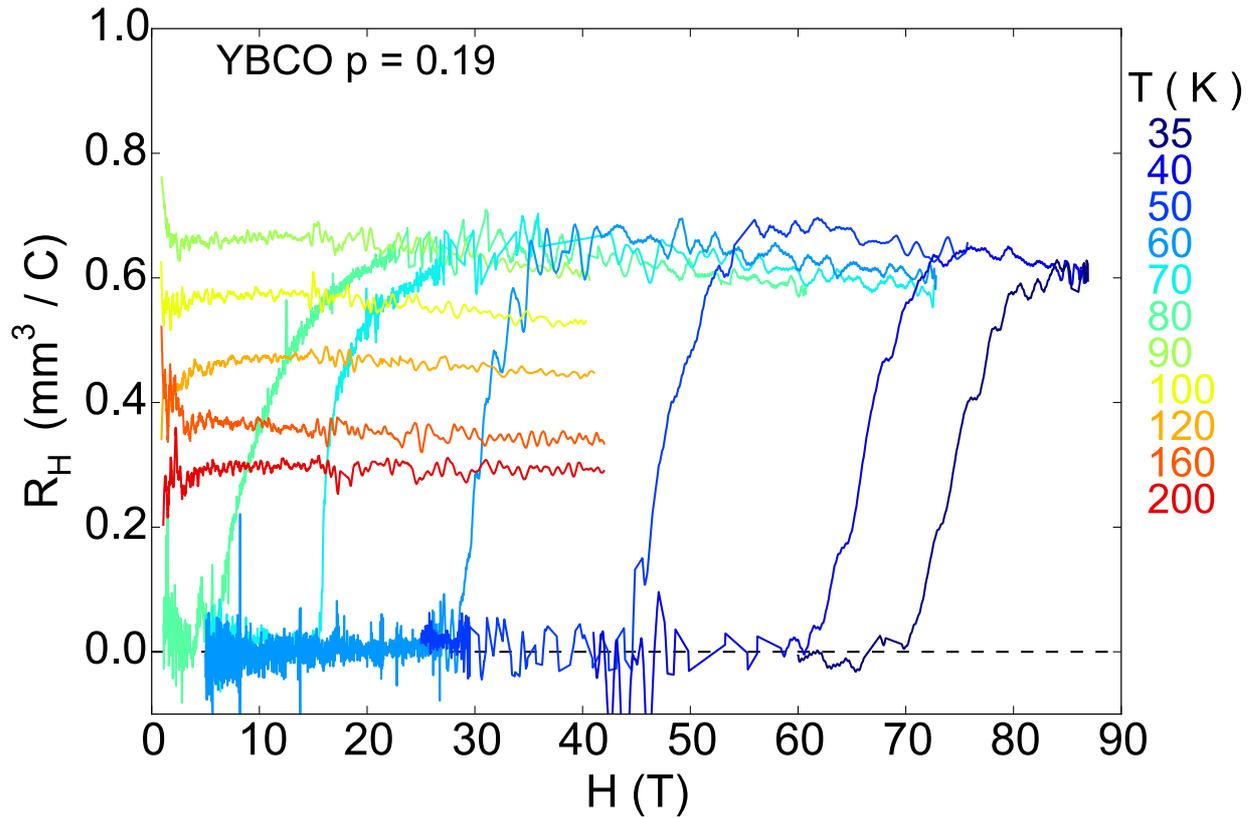

**Figure S3 | Isotherms of $R_H$ vs $H$ in YBCO at $p$ = 0.19.**

Magnetic field dependence of the Hall coefficient $R_H$ in our YBCO sample with $y$ = 6.99 and 1.4 % Ca doping ($T_c$ = 87 K; $p$ = 0.19), at various temperatures as indicated.

**Supplementary Figure S4**

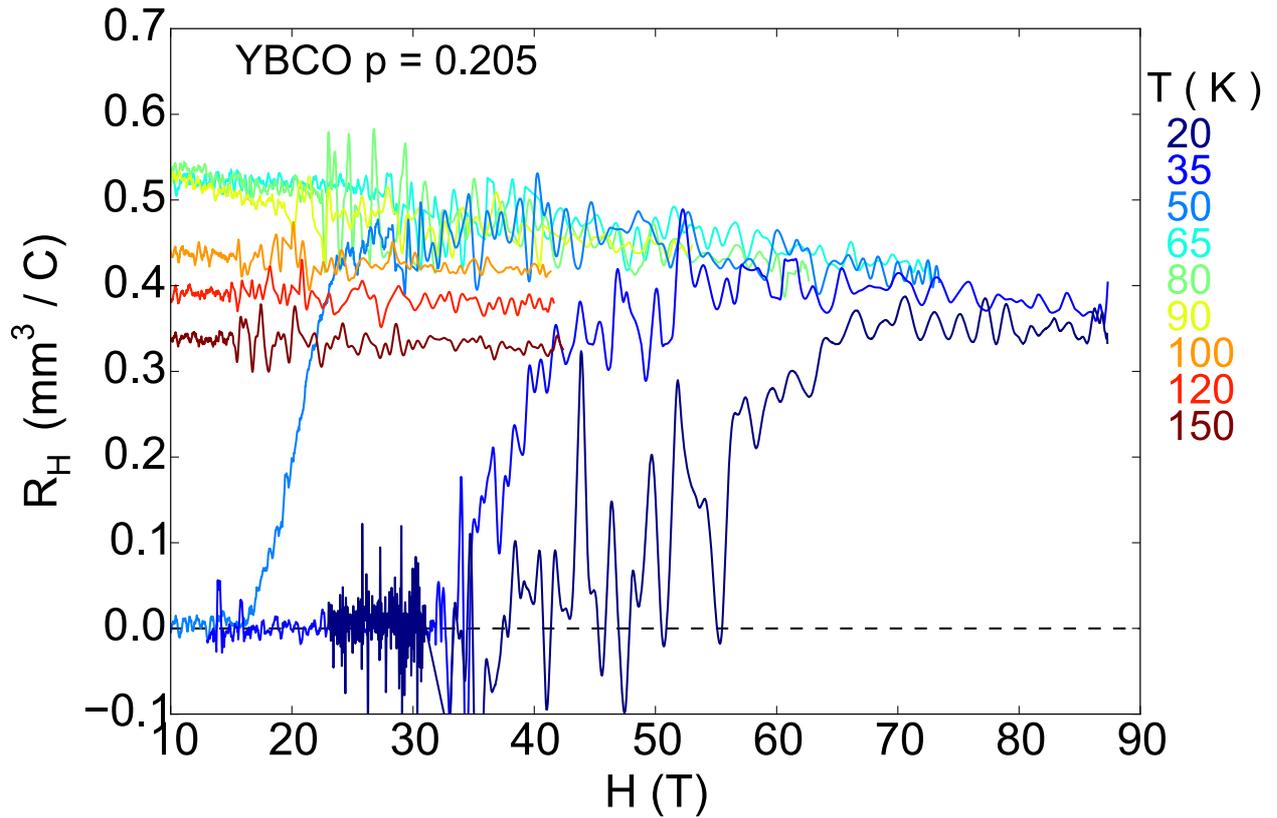

**Figure S4 | Isotherms of $R_H$ vs $H$ in YBCO at $p$ = 0.205.**

Magnetic field dependence of the Hall coefficient $R_H$ in our YBCO sample with $y$ = 6.99 and 5 % Ca doping ($T_c$ = 77 K; $p$ = 0.205), at various temperatures as indicated.

**Supplementary Figure S5**

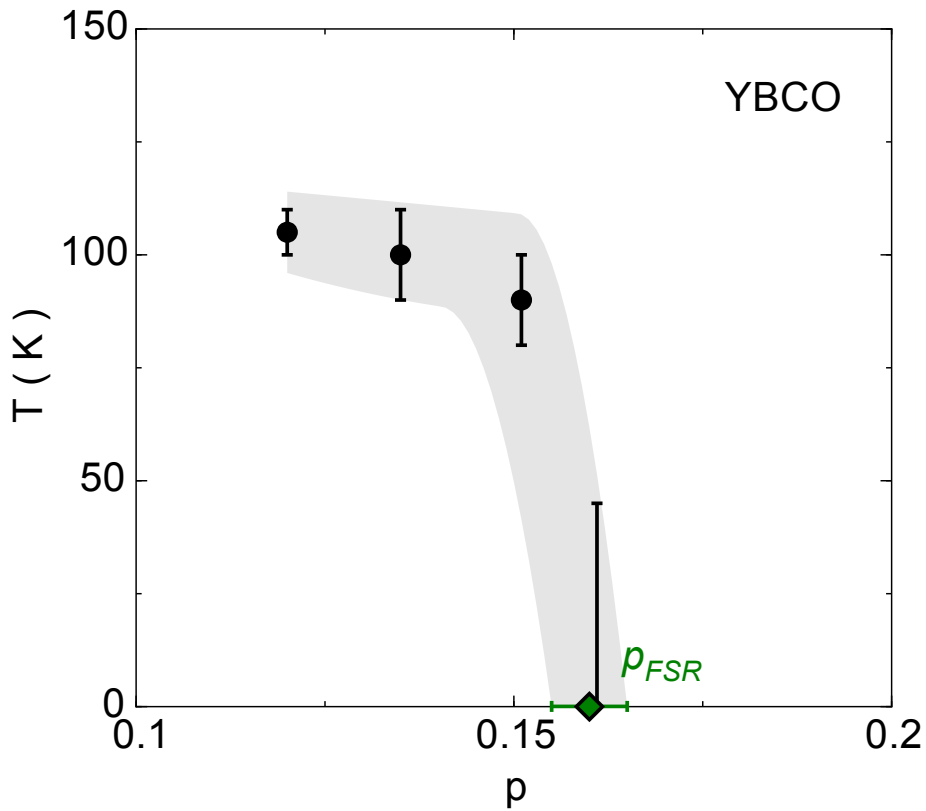

**Figure S5 | Doping dependence of $T_{max}$.**

Temperature $T_{max}$ at which $R_H$ vs $T$ peaks in YBCO (Fig. 3a), plotted vs doping $p$. At $p = 0.16$, there is no downturn in the normal-state $R_H(T)$ down to 40 K. The $p = 0.16$ data are consistent with $T_{max} = 0$ (lower bound), with an upper bound at $T_{max} = 40$ K. The width of the grey band marks the upper and lower limits for $T_{max}$ vs $p$. The green diamond defines the critical doping above which FSR is no longer present, at $p_{FSR} = 0.16 \pm 0.005$.

**Supplementary Figure S6**

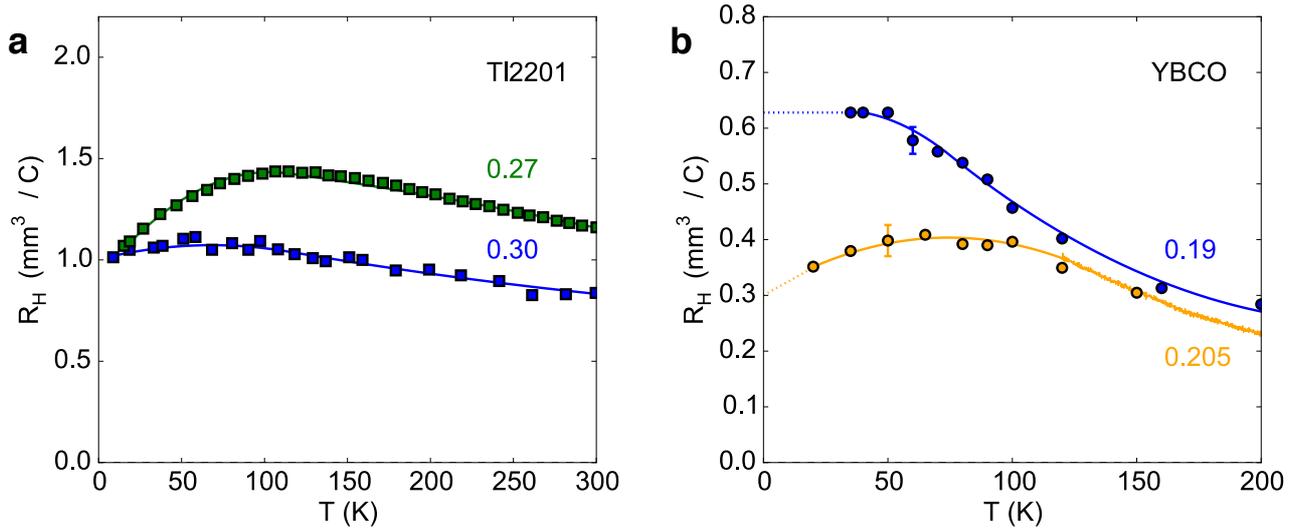

**Figure S6 | Zoom on $R_H$ vs $T$ in Tl-2201 and YBCO at high doping.**

**a)** Temperature dependence of $R_H$ in Tl-2201 (squares) at $p = 0.3$ (blue, $T_c = 10$ K; ref. 35) and $p = 0.27$ (green, $T_c = 25$ K; ref. 36). **b)** $R_H$ vs $T$ in YBCO (circles, from Figs. S3, S4 and S7) at $p = 0.205$ (yellow) and $p = 0.19$ (blue). The dotted lines are an extrapolation of the low-$T$ data to $T = 0$. The YBCO curve at $p = 0.205$ is qualitatively similar to the two Tl-2201 curves, all exhibiting an initial rise with increasing temperature from $T = 0$, and a characteristic peak at $T \sim 100$ K – two features attributed to inelastic scattering on a large hole-like Fermi surface [15]. The YBCO curve at $p = 0.19$ is qualitatively different, showing no sign of a drop at low $T$ (see Fig. S7). We attribute the two-fold increase in the magnitude of $R_H$ at $T \rightarrow 0$ to a decrease in carrier density as the pseudogap opens at $p^*$, with $p^*$ located between $p = 0.205$ and $p = 0.19$.

**Supplementary Figure S7**

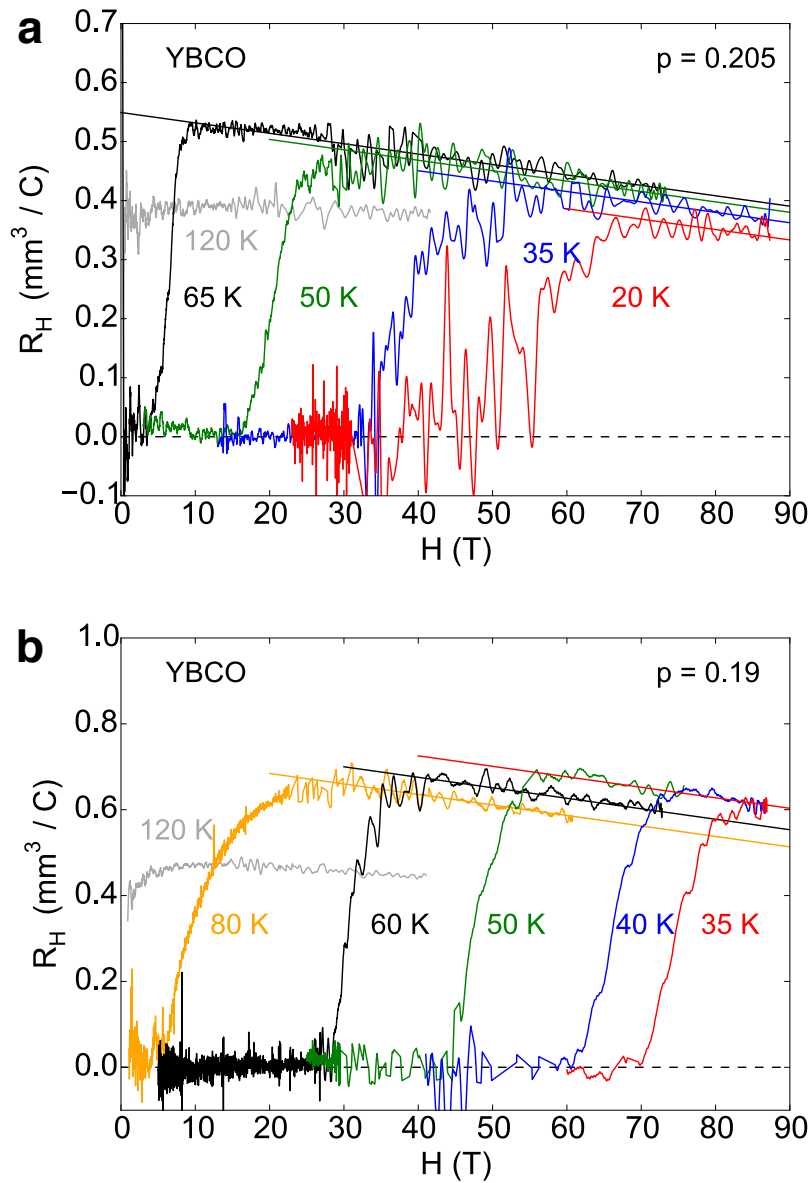

**Figure S7 | Comparison between $p$ = 0.205 and $p$ = 0.19.**

Field dependence of the Hall coefficient $R_H$ in YBCO at **a)** $p$ = 0.205 and **b)** $p$ = 0.19, for different temperatures as indicated. The color-coded lines are parallel linear fits to the high-field data. They show that at low temperature $R_H$ decreases upon cooling at $p$ = 0.205, while it saturates at $p$ = 0.19. The value of $R_H$ given by the fit line, at $H$ = 80 T, is plotted in Fig. 3 and in Fig. S6b. Similar fits are used to extract $R_H$(80 T) for $p$ = 0.16 and $p$ = 0.177 (from data in Figs. S1 and S2).

**Supplementary Figure S8**

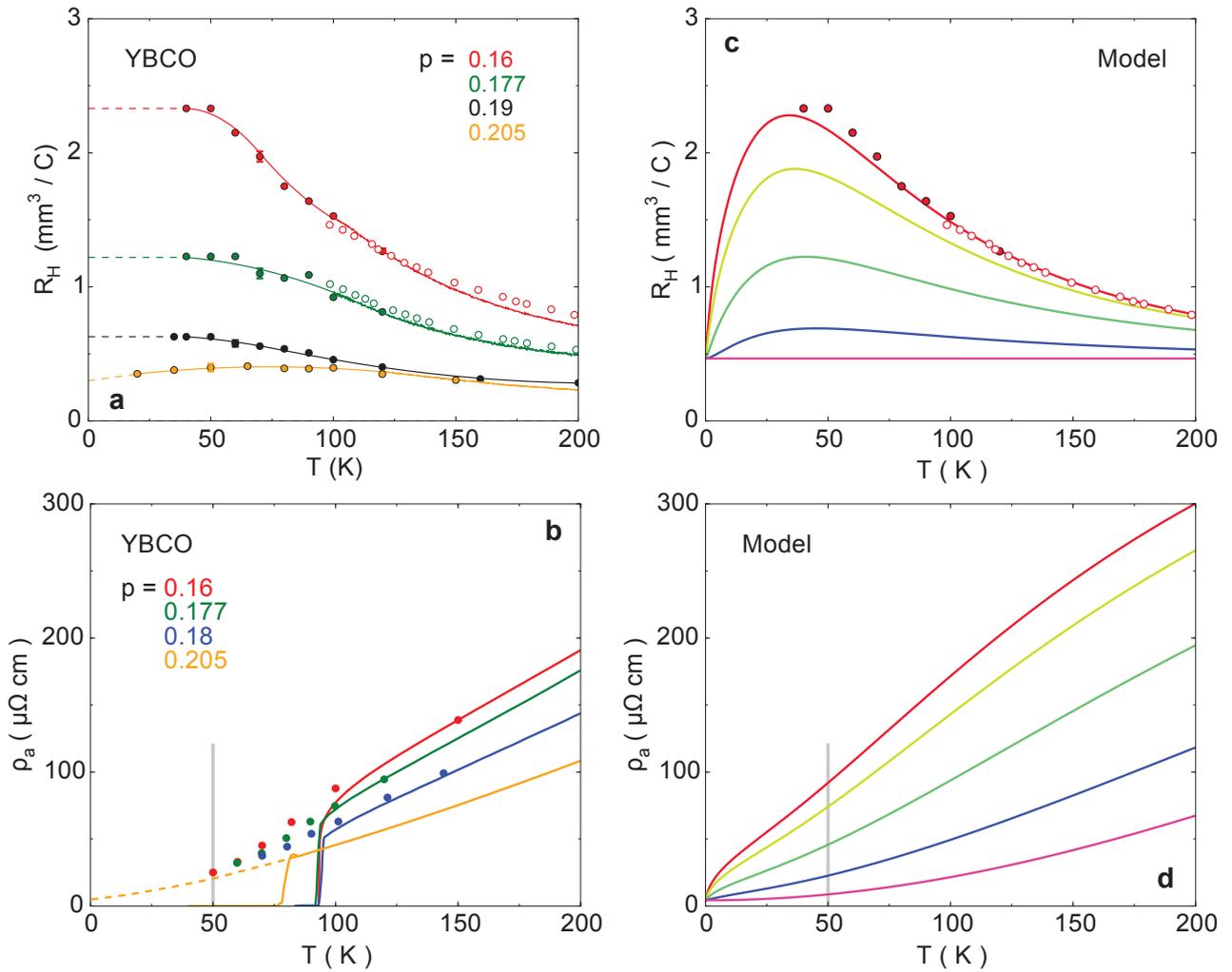

**Figure S8 | Scenario of inelastic scattering.**

a) $R_H$ vs $T$ in YBCO at 4 dopings, as indicated (Fig. 3b). b) Electrical resistivity $\rho_a$ vs $T$ in YBCO at 4 dopings, as indicated. Lines are at $H = 0$; dots are in the normal state at high field. c) $R_H$ vs $T$ calculated for 5 values of inelastic scattering, with $\Gamma_1 = 0, 1, 5, 15$ and $25$ THz / K, showing that $R_H(T)$ grows with increasing $\Gamma_1$ (see Supplementary discussion). Dots are from panel a. d) Corresponding calculated values of the electrical resistivity $\rho_a$, plotted vs $T$, using the same parameters and values of $\Gamma_1$ as for the color-coded curves of panel c. The vertical grey lines mark $T = 50$ K, the temperature at which we see a 6-fold increase in $R_H$ (a), yet no increase in $\rho_a$ (b). The calculation can reproduce the large increase in $R_H$ (c), but it is accompanied by a 10-fold increase in $\rho_a$ (d).

# Methods

## SAMPLES

Single crystals of YBa$_2$Cu$_3$O$_y$ (YBCO) were obtained by flux growth at UBC [37]. The superconducting transition temperature $T_c$ was determined as the temperature below which the zero-field resistance $R = 0$. The hole doping $p$ is obtained from $T_c$ [38]. In order to access dopings above $p = 0.18$, Ca substitution was used, at the level of 1.4 % (giving $p = 0.19$) and 5 % (giving $p = 0.205$). The samples are rectangular platelets with six contacts applied in the standard geometry, using diffused gold pads.

## MEASUREMENT OF THE LONGITUDINAL AND TRANSVERSE RESISTANCES

The longitudinal resistance $R_{xx}$ and transverse (Hall) resistance $R_{xy}$ of our YBCO samples were measured in Sherbrooke in steady fields up to 16 T and in Toulouse in pulsed fields up to 88 T, using a dual coil magnet developed at the LNCMI-Toulouse to produce non-destructive magnetic fields up to 90 T. The magnetic field profile is shown in Fig. S9.

The pulsed-field measurements were performed using a conventional 6-point configuration with a current excitation between 5 mA and 10 mA at a frequency of ~ 60 kHz. A high-speed acquisition system was used to digitize the reference signal (current) and the voltage drop across the sample at a frequency of 500 kHz. The data were post-analyzed with a software to perform the phase comparison. Data for the rise and fall of the field pulse were in good agreement, thus excluding any heating due to eddy currents. Tests at different frequencies showed excellent reproducibility.

## ERROR BARS

Note that the resistance of the samples was small due to their geometric factor and their high conductivity in this doping range – typically a few milliohms in the normal state at high fields. Despite the fact that $R_{xy}$ was obtained by anti-symmetrizing the signals measured for a field parallel and anti-parallel to the $c$ axis, a slight negative slope was observed in the Hall coefficient $R_H$ vs $H$, similar to that found in prior high-field studies [17, 21]. This slope, which may be intrinsic or not, has no impact on any of our conclusions, since they do not depend on the precise absolute value of $R_H$. Indeed, our conclusions depend on two results: 1) the temperature dependence of $R_H$ at low *T*, in a given sample; 2) the doping dependence of $R_H$ at low *T*, at a given temperature. In both cases, what matters is to measure $R_H$ at the same value of *H*, namely *H* = 80 T. So in Figs. 3c, S6b and S7, where we compare the detailed temperature dependence of $R_H(T)$ in two samples (*p* = 0.19 and *p* = 0.205), the relevant uncertainty is the relative error bar associated with a change of temperature in one sample. That error is defined as the standard deviation in the value of $R_H$ at *H* = 80 T given by the linear fit in Fig. S7. The maximum such error bar is shown in Fig. 3 for each of our four samples.

In Fig. 4a, we simply compare the magnitude of $R_H$ in our four samples when measured at *H* = 80 T and *T* = 50 K. As can be seen from the raw data, the negative slope of $R_H$ vs *H* does not really affect this comparison. What comes in is the error bar on the absolute value of $R_H$ (in mm$^3$/C), which involves geometric factors and which we estimate to be at most ± 15 %. This error bar is shown in Fig. 4b. Note the excellent quantitative agreement between our data and the data of ref. 16 at *p* = 0.16 and 0.177 (Fig. 3b).

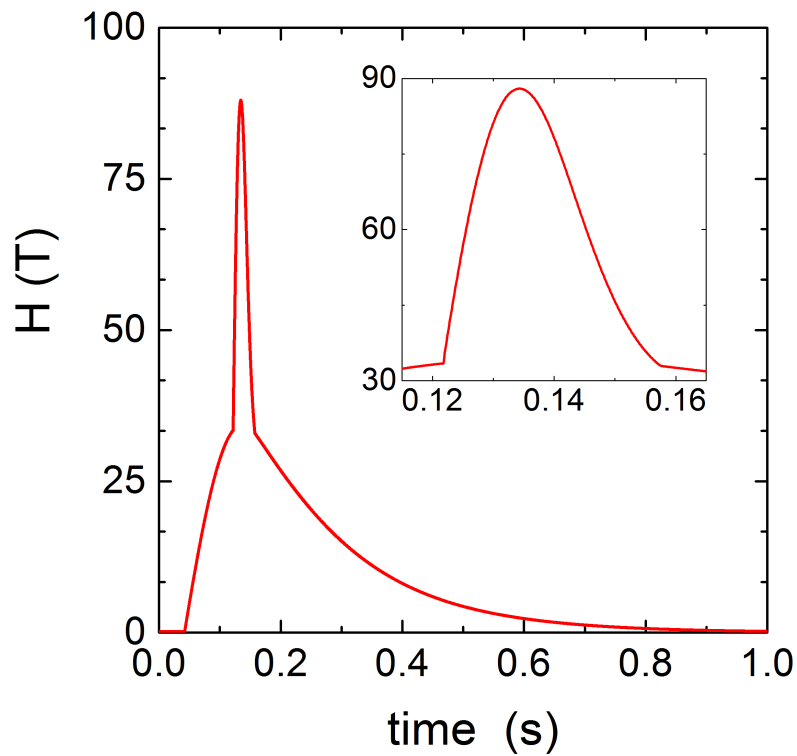

**Figure S9 | Magnetic field profile.** Time dependence of the magnetic field pulse in the 90 T dual-coil magnet at the LNCMI in Toulouse. <u>Inset</u>: zoom around maximum field.

## Supplementary Discussion

**CALCULATION OF HALL COEFFICIENT AND RESISTIVITY IN CUPRATES**

Assuming a single large hole-like Fermi surface, as measured in strongly overdoped Tl-2201, Hussey has shown that one can calculate the resistivity and Hall coefficient using the Jones-Zener expansion [39]. The model calculates directly the longitudinal and transverse electrical conductivities $\sigma_{xx}$ and $\sigma_{xy}$:

$$\sigma_{xx} = \frac{e^2}{4\pi^3 \hbar}\left(\frac{2\pi}{d}\right) 4 \int_0^{\pi/2} \frac{k_F v_F \cos^2(\varphi - \gamma)}{\Gamma \cos \gamma} d\varphi$$

$$\sigma_{xy} = \frac{-e^3 H}{4\pi^3 \hbar^2}\left(\frac{2\pi}{d}\right) 4 \int_0^{\pi/2} \frac{v_F \cos(\varphi - \gamma)}{\Gamma} \frac{\partial}{\partial \varphi}\left(\frac{v_F \sin(\varphi - \gamma)}{\Gamma}\right) d\varphi$$

Therefore:

$$R_H = \frac{\sigma_{xy}}{\sigma_{xx}^2 + \sigma_{xy}^2} \frac{1}{H}$$

$$\rho_{xx} = \frac{\sigma_{xx}}{\sigma_{xx}^2 + \sigma_{xy}^2}$$

with $e$ the electron charge, $\hbar$ the Plank constant, $d$ distance between two CuO$_2$ planes, $k_F$ the Fermi momentum, $v_F$ the Fermi velocity, $\varphi$ is the angle between the momentum **k** and $k_x$ axis in the first Brillouin zone (FBZ), $\gamma(\varphi) = \tan^{-1}\left[\frac{\partial}{\partial \varphi}(\log k_F(\varphi))\right]$, and $\Gamma$ the scattering rate (later named $\Gamma_{eff}$. Here we choose $k_F$ and $v_F$ not to be $\varphi$-dependent, i.e. the Fermi surface is a perfect cylinder, implying $\gamma(\varphi) = 0$.

We calculate $k_F$ and $v_F$ from hole doping $p$ and effective mass $m^*$ (= 4.1 $m_e$ from quantum oscillations observed in overdoped Tl-2201 [40] ):

$$n = \frac{1+p}{a^2}$$

$$k_F = \sqrt{\frac{2 e n \Phi_0}{\hbar}}$$

$$v_F = \frac{\sqrt{2\,e\,\hbar\,n\,\Phi_0}}{m^*}$$

where $a$ is the in-plane lattice constant parameter (we neglect the slight orthorhombicity of YBCO), $n$ the carrier density per $CuO_2$ plane, and $\Phi_0$ the flux quantum.

## SCENARIO OF INELASTIC SCATTERING APPLIED TO YBCO

Here we discuss the possibility that $R_H$ in YBCO at low temperature is enhanced not by a loss of carrier density but by an increase in inelastic scattering.

It has been shown that anisotropic inelastic scattering can increase the value of $R_H(T)$ even if the Fermi surface remains a single large isotropic cylinder [15, 39]. This mechanism has been argued to account for the rise in $R_H$ measured in overdoped Tl-2201, as occurs when the doping is decreased from $p$ = 0.3 to $p$ = 0.27, for example (Fig. S6a).

Here we use the following inelastic scattering model developed by Hussey [15], where the effective scattering rate is given by:

$$1/\Gamma_{eff}(T, \varphi) = 1/\Gamma_{ideal} + 1/\Gamma_{max}$$

with $\Gamma_{ideal}(T, \varphi) = \Gamma_0 + \Gamma_1 \cos^2(2\varphi)\, T + \Gamma_2\, T^2$ and $\Gamma_{max} = v_F / a$ ; where $T$ is the temperature, $\Gamma_0$ is the elastic rate scattering coefficient, $\Gamma_1$ is the $T$-linear inelastic scattering rate coefficient, $\Gamma_2$ is the $T^2$ scattering rate, $\Gamma_{max}$ is the maximum scattering rate limited by the lattice constant $a$.

Here we use this model to fit our Hall data for YBCO at $p$ = 0.16, with $\Gamma_1$ and $\Gamma_2$ the only free parameters ($\Gamma_0$ is chosen so that the calculated value of $\rho_{xx}$ at $T$ = 0 agrees with experiment). The resulting fit is shown in Fig. S8c (solid red line). The corresponding curve of $\rho_{xx}(T) = \rho_a(T)$ is plotted in Fig. S8d (solid red line).

In Fig. S8, we show how these calculated curves vary when the strength of inelastic scattering is varied, both on $R_H$ (Fig. S8c) and on $\rho_a$ (Fig. S8d). The calculated curves may be compared with experimental curves in YBCO, shown in the left panels of Fig. S8, namely $R_H$ vs $T$ in Fig. S8a and $\rho_a$ vs $T$ in Fig. S8b. We see that by choosing a large value of $\Gamma_1$, one can fit the data at $p$ = 0.16 quite well. The calculated curve drops precipitously below the lowest experimental data point. The decrease in the overall magnitude of $R_H$ vs $T$ with doping can be mimicked in the calculations by decreasing $\Gamma_1$ gradually to zero, at which point $R_H$ becomes constant.

However, while the calculated curves are consistent with the measured $R_H$, they are in complete disagreement with the measured $\rho_{xx} = \rho_a$. This is seen by comparing calculated (Fig. S8d) and measured (Fig. S8b) values. We see that the 10-fold increase in the calculated $\rho_a$ at 50 K, caused by the large increase in $\Gamma_1$, is not at all observed in the experimental data, which are essentially independent of doping at $T$ = 50 K. In other words, if inelastic scattering were responsible for the increase in $R_H$ at 50 K with underdoping, it would necessarily show up as a comparable (even larger) increase in the resistivity $\rho_a$ at 50 K. The fact that it doesn't rules out inelastic scattering as a mechanism for the 6-fold increase in $R_H$.

We conclude that the large rise in $R_H$ vs doping is due to a loss of carrier density, and it is a property of the normal-state Fermi surface at $T$ = 0.